\documentclass[aip,pof,onecolumn]{revtex4-1}

\pdfoutput=1

\usepackage{graphicx}
\usepackage[applemac]{inputenc}
\usepackage{natbib}
\usepackage{bm}
\usepackage{amssymb}
\usepackage{amsbsy}
\usepackage[usenames]{color}
\usepackage{mathrsfs,mathcomp}
 \usepackage[normalem]{ulem}
\usepackage{amsmath}
\usepackage{amsfonts}
\usepackage{mathrsfs}
\usepackage{subfigure}
\usepackage{ifsym} 
\usepackage{multirow}
\usepackage[margin=1.1in]{geometry}

\begin{document}

\title{Short time dynamics of viscous drop spreading}
\author{A. Eddi}
 \email{antonin.eddi@utwente.nl}
\author{K.G. Winkels}%
\author{J.H. Snoeijer}
\affiliation{ Physics of Fluids Group and MESA+ Institute for Nanotechnology, Faculty of Science and Technology, University of Twente, 7500AE Enschede, The Netherlands}

\date{\today}

\begin{abstract}
Liquid drops start spreading directly after coming into contact with a solid substrate. Although this phenomenon involves a three-phase contact line, the spreading motion can be very fast. We experimentally study the initial spreading dynamics, characterized by the radius of the wetted area, for viscous drops. 
Using high-speed imaging with synchronized bottom and side views gives access to 6 decades of time resolution. We show that short time spreading does not exhibit a pure power-law growth. Instead, we find a spreading velocity that decreases logarithmically in time, with a dynamics identical to that of coalescing viscous drops. Remarkably, the contact line dissipation and wetting effects turn out to be unimportant during the initial stages of drop spreading.
\end{abstract}
\pacs{47 55.D- Drops}
\maketitle

\section{Introduction}

\begin{figure}[t]
\includegraphics[width=1 \columnwidth]{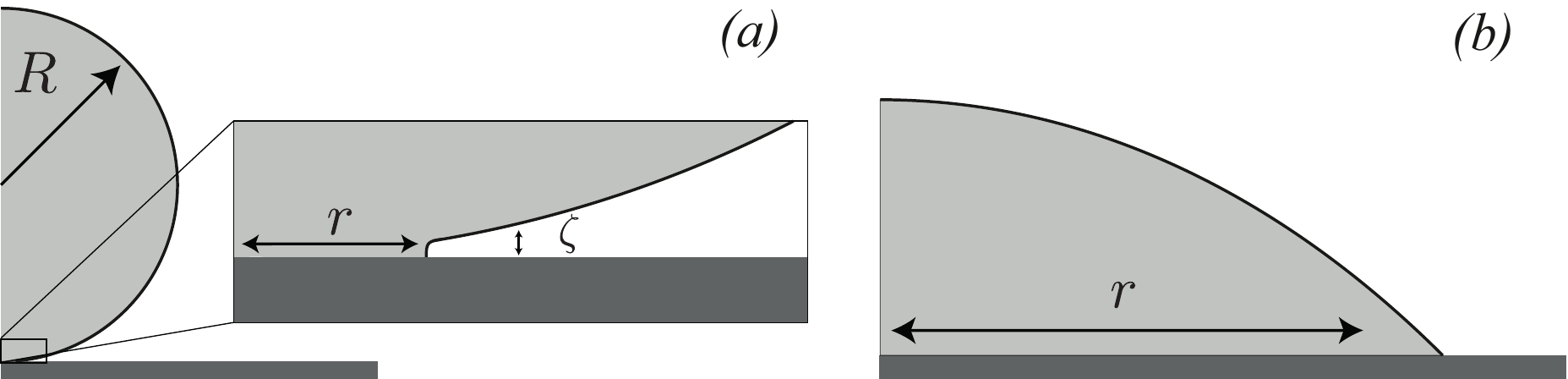}
\caption{\label{fig1} (a) Schematics presenting the geometry of a drop with initial radius $R$ during the initial stages of spreading. The width of the narrow gap $\zeta$ determines the local curvature and thus the driving force. It depends on the radius of the wetted area $r$. (b) Schematics showing the geometry of the drop during the late stages of spreading.}
\end{figure}

Liquid drops start spreading directly after coming into contact with a solid substrate. This process is relevant for many applications, from printing and coating to agriculture \cite{Wijshoff:2010,Simpkins:2003,Bonn:2009,Vovelle:2000,deGennes:1985}. More precisely, we will consider here a drop of initial radius $R$ that is gently brought into contact with a flat substrate (see fig. \ref{fig1}). The liquid has a surface tension $\gamma$, a viscosity $\eta$ and a density $\rho$, while the interaction with the substrate is characterized by the equilibrium contact angle $\theta_{eq}$. During spreading, the contact line moves radially outwards from the contact point and the drop wets a circular area of radius, $r(t)$. For the late times of spreading, the drop assumes the shape of a spherical cap (see fig. \ref{fig1}(b)). The dynamics is governed by viscous effects near the contact line and characterized an apparent (dynamic) contact angle. When the substrate is completely wetting, this is described by the so-called Tanner's law \cite{Tanner:1979,deGennes:2003,Bonn:2002}

\begin{equation}
\frac{r}{R} \sim \left(\frac{\gamma t}{\eta R}\right)^{1/10} ~.
\label{Tanner}
\end{equation}
It is well-known that the flow near the moving contact line is singular, since the viscous stresses diverge in a liquid wedge with a no-slip boundary condition \cite{Huh:1971}. This singularity is regularized by introducing a cut-off length at molecular scales to avoid the divergence. Once regularized, the visco-capillary balance near the contact line gives rise to the Cox-Voinov law for the apparent contact angle \cite{Voinov:1976,Cox:1986}. Combined with the spherical cap geometry of the drop, one naturally obtains  Eq.~(\ref{Tanner}).

While Tanner's law predicts the late time behavior of spreading for viscous drops, it does not apply at short times. In that case the geometry of the drop is completely different (fig. \ref{fig1}(a)). Just after contact, the drop shape is highly curved close to the contact point, with a meniscus characterized by a small length $\zeta$ as indicated in the sketch. This strongly curved meniscus generates a rapid flow inside the drop. In fact, the initial contact represents a singularity in itself: the meniscus size $\zeta\rightarrow 0$ as $r \rightarrow 0$, inducing diverging capillary stresses. 

Interestingly, the geometry in fig. \ref{fig1}(a) is strongly reminiscent to the coalescence of two spherical drops, if we consider the substrate to act as a mirror-plane for the flow. This analogy was first employed by Biance et al.~\cite{Biance:2004}, who compared the spreading to the well-studied growth of the neck when two identical drops coalesce \cite{Aarts:2005,Wu:2004,Thoroddsen:2005,Case:2008,Paulsen:2011}. This approach has proven very successful in the low-viscosity limit. The coalescence of water drops \cite{Aarts:2005} shows the same dynamics as water drops spreading in total wetting conditions \cite{Biance:2004,Bird:2008,Courbin:2009,Winkels:2012}. In both cases it was found that $r \sim t^{1/2}$, with prefactors in agreement with potential flow calculations \cite{Eggers:2003}.

A natural question that arises is whether the spreading-coalescence analogy also holds for drops of high viscosity. In this limit, coalescence experiments \cite{Paulsen:2011} are in good agreement with predictions based on Stokes flow calculations \cite{Hopper:1990, Hopper:1993,Eggers:1999}. The driving force depends on the radius of curvature $\zeta$ in the neck, and the speed $u$ of the neck is given by \cite{Eggers:1999}

\begin{equation}
u=\frac{dr}{dt}\simeq -\frac{1}{2\pi} \frac{\gamma}{\eta} \ln{\frac{\zeta}{r}}~.
\label{eq_speed}
\end{equation}
This result is essentially the Stokeslet solution in a two-dimensional flow, where the meniscus gap $\zeta$ provides the inner cutoff length for the point force. It was shown analytically that $\zeta\sim r^3/R^2$ when the surrounding fluid has no viscosity \cite{ Hopper:1993}, while $\zeta\sim r^{3/2}/R^{1/2}$ when the outer fluid is viscous \cite{Eggers:1999}. In the latter case, the radius of the neck in between the two drops is described asymptotically by

\begin{equation}
{r}\simeq -~\frac{1}{4 \pi}~ \frac{\gamma}{\eta} ~t~ \ln{\frac{r}{R} }~.
\label{eq_radius}
\end{equation}
Hence, the neck radius $r$ grows linearly in time with logarithmic corrections. The same law is obtained when the outer fluid viscosity is negligible, the numerical prefactor being 4 times larger.

In this paper we present an experimental investigation of viscous drop spreading on a substrate. In Sec. \ref{section2}, we first introduce the experimental set-up that allows for an increased resolution both in time and space with respect to previous experiments. In Sec. \ref{section3}, we discuss the qualitative and quantitative nature of this improvement. This is a crucial step in order to access a sufficient number of decades to reveal the initial spreading dynamics. Section \ref{section4} summarizes the experimental results obtained for water-glycerin mixtures. We investigate the influence of the initial drop radius $R$, the equilibrium contact angle $\theta_{eq}$, and the viscosity $\eta$. Our key findings are that the viscous spreading dynamics is not a simple power-law of the type $r \sim t^\alpha$, and that the initial stages of spreading are independent of the substrate wettability. Instead, the spreading is consistent with the coalescence-prediction given in Eq. (\ref{eq_radius}). In Sec. \ref{section5}, we further discuss our findings in the light of this comparison between drop coalescence and drop spreading.

\section{Experimental set-up}
\label{section2}
\begin{table}[t]
\begin{centering}
\renewcommand{\arraystretch}{1.8}
\begin{tabular}{|c|c|c|c|}
\hline
~~Mixture ~~&$~~\eta$ (mPa.s)~~ &~~ $\gamma$ (mN.m$^{-1}$)~~ & $~~\rho$ (kg.m$^{-3}$) ~~\\ \hline
~Pure Glycerine ~ & $1120\pm 30$ & $63.1\pm 0.2$  & $1262 \pm 5$ \\ \hline
~Glycerine/Water -- $90/10 \%$~ & $220 \pm 5$ & $63.4 \pm 0.2$  & $1238 \pm 5$ \\ \hline
~Glycerine/Water -- $79/21 \%$~ & $50 \pm 2$  & $64.7 \pm 0.2$ & $1204 \pm 5$ \\ \hline
~Glycerine/Water -- $60/40 \%$~ & $11.5 \pm .5$ & $ 67.3 \pm 0.2$ & $1153 \pm 5$  \\ \hline
\end{tabular}
\end{centering}
\caption{\label{table1} Properties of the different water-glycerine mixtures used in the experiment}
\end{table}
\begin{figure}[t]
\includegraphics[width=1 \columnwidth]{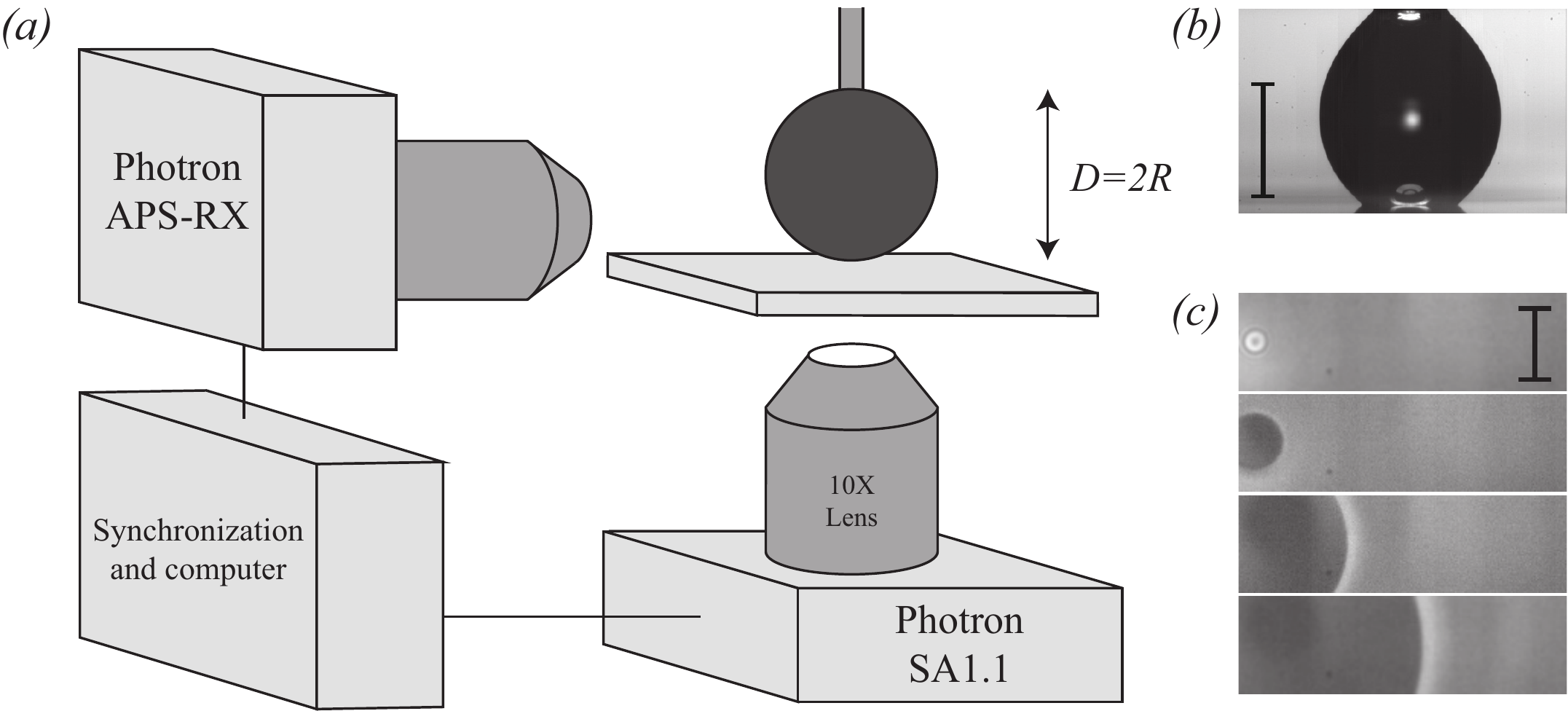}
\caption{\label{fig2} (a) Sketch of the experimental set-up. A drop of water-glycerine mixture attached to the tip of a needle is brought in contact with a glass substrate. Two high-speed cameras allow for synchronized bottom view and side view recordings. (b) Typical side view image of a spreading drop (the black bar indicates $1$ mm). (c) Series of bottom view snapshots of a spreading drop, on the last frame before contact and at times $9\cdot10^{-5}$~s, $1.7\cdot10^{-4}~$s and $4.9\cdot10^{-4}~$s after contact (the black bar indicates $100~\mu$m). Prior to contact, we observe an interference pattern at the location where the drop will touch the substrate. After contact, we observe the growth of a circular dark area corresponding to the wetted area on the glass substrate.}
\end{figure}

The initial stages of drop spreading are investigated using the set-up shown in fig. \ref{fig2}. A drop of liquid is created at the tip of a needle and inflated quasi-statically. The needle is fixed at a distance $D$ above a glass slide, so that the drop touches the substrate when it reaches a radius $R=D/2$. The fluid is injected using a syringe pump Picoplus (Harvard Apparatus, USA) set at a minimal outflux. We check that the approach speed at contact is lower than 10 $\mu$m.s$^{-1}$, eliminating any initial dynamics. We use water-glycerin mixtures in order to vary the viscosity of the liquid. The viscosity ranges from $\eta=11.5$ to $1120$ mPa.s. Surface tensions and contact angles are measured using a OCA 15PRO tensiometer (Dataphysics, Germany). Viscosity is measured using a Rheolab QC rheometer (Anton Parr, Austria) and density using a DMA 35 densimeter (Anton Parr, Austria). Physical properties of the various water-glycerine mixtures are summarized in table \ref{table1}. 

In most experiments, we use glass substrates that are completely wetted by the liquids. Microscope glass slides (Menzel, Germany) are cleaned using ethanol and acetone, then submerged in an ultrasonic bath for 15 mn and dried with filtered nitrogen gas. For partial wetting experiments, we coated glass substrates. The advancing contact angles for these substrates are presented in section \ref{section4}.B.

We use two synchronized high-speed cameras to image the spreading dynamics of the drop (see fig. \ref{fig2}). An APX-RS camera coupled to a long-range microscope (Navitar 12X Zoom coupled with a 2X adapter tube) allows for a side view of the spreading drop (see fig. \ref{fig2}(b)). The spatial resolution is $3.5~\mu$m/pixel, and the acquisition rate is 10 000 frames/s. A Photron SA1.1 camera coupled to an inverted microscope (Zeiss Axiovert 25, objective Zeiss A-plan 10 $\times$ magnification) records a bottom view of the wetted area (see fig. \ref{fig2}(c)). This allows for high speed recording of the spreading, up to 250 000 frames/s, with a spatial resolution of $2~ \mu$m/pixel. Before contact, the drop is pendant a few microns above the glass surface, and we can observe circular fringes. After contact, we see a circular dark area (the wetted surface) growing with time. Contact occurs between the last frame where fringes can be seen and the first frame where the dark area appears. In order to minimize the error on the contact time, we set $t=0$ half way between these two recorded frames. 

The recorded images are processed using custom Matlab routines in which the position of the contact line of side and bottom view images is detected using a convolution procedure \cite{Marchand:these}. The contact line is located at the place were the intensity profile (in gray scale levels) presents its maximum slope. We repeated each experiments several times. Results turn out to be extremely reproducible and the main uncertainty at short times results from spatial resolution rather than from differences between successive experiments. The error bars given in the plots correspond to an uncertainty of 1 pixel in the detected radius.

In summary, this set-up gives access to 6 decades of time resolution, from a few microseconds up to a few seconds, and 3 decades of space resolution, from a few microns up to a few millimeters.

\section{Side view and bottom view}
\label{section3}

Most of the spreading experiments in which high speed imaging was used where done using only side view recordings \cite{Biance:2004, Bird:2008, Chen:2011, Carlson:2011, Carlson:2012}. We recently introduced a bottom view imaging that  highly enhances the spatial resolution and thus significantly improves the temporal resolution of such experiments \cite{Winkels:2012}. When looking at previous experimental studies based on side view measurements, all cases exhibit a plateau at short times corresponding to a non vanishing wetted radius at contact. Here, we would like to discuss the origin of this plateau.

\begin{figure}[t]
\begin{centering}
\includegraphics[width=1 \columnwidth]{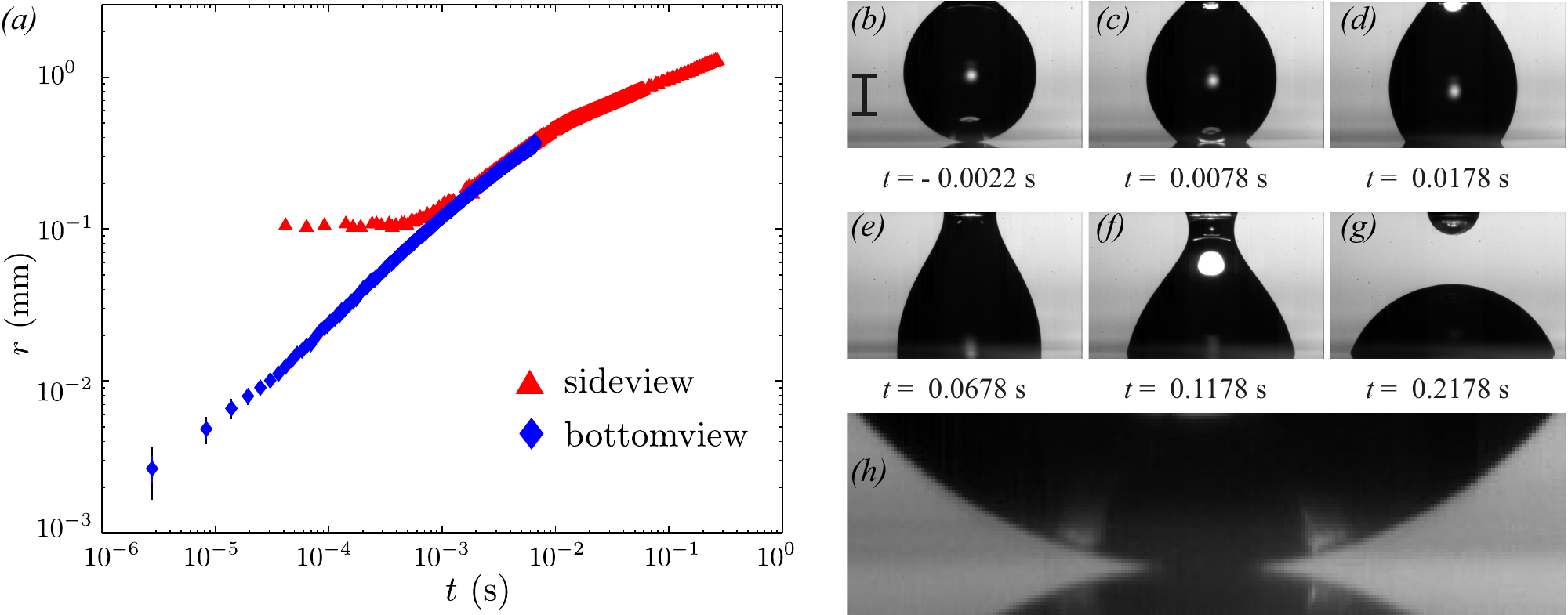}
\caption{\label{fig3} Evolution of the spreading radius $r$ as a function of time. The radius is simultaneously determined from the bottom view (blue diamonds) and from the side view (red triangles). Both  measurements agree at times larger than $10^{-3}$~s, but there is a large discrepancy at shorter times. The side view data present a horizontal plateau whereas the bottom view data suggest that $r$ vanishes as $t\rightarrow0$. (b-g) Series of side view snapshots of a spreading drop. The black bar represents 0.5 mm. (h) Magnified side view of the drop at the last frame before contact ($t=-0.0022~$s). The drop and its mirror image in the glass plate are already merged. This leads to a non-zero value of $r$ before contact, and thus to an over-estimation of the contact radius at short times.}
\end{centering}
\end{figure}
Fig. \ref{fig2} shows the radius $r$ of the wetted area for a drop of water-glycerin mixture with viscosity $\eta=220~$mPa.s. The blue squares indicate the bottom view and red triangles the synchronized side view. We see that, for times $t<10^{-3}~$s, measurements disagree, so that the side view present the same plateau as in former studies, whereas the bottom view  suggests that $r$ vanishes when $t\rightarrow 0$. It is only later, at times larger than $10^{-3}~$s that the two curves collapse.

The discrepancy can be understood as follows. Fig. \ref{fig3} presents a series of side view images extracted from this recording. It is important to note that the first image corresponds to a time before the actual contact between the drop and the substrate (determined from the bottom view). Fig. \ref{fig3} (g) is a magnification of this image around the contact point. Even though there is no contact at that time (we still see fringes on the corresponding bottom view images), it looks as if the drop already touches the substrate. Due to the presence of a narrow gap and optical limitations, the drop image merges with its own optical reflexion on the glass substrate. This effect is enhanced by the circular shape of the drop: only a few pixels of merging in the vertical direction dramatically increases the measured radius $r$.  All these reasons lead to an overestimation of $r$ and thus to the observed plateau. It is only at later times, when the radius $r$ becomes larger than a few hundreds of microns (being of the same order as the macroscopic radius of the drop) that this optical effect can be neglected and that the side view measurements give identical results as the bottom view. The bottom view does not suffer from this artifact: both $t=0$ and $r$ can be determined accurately. Hence, all our experimental results will be based upon bottom view measurements for radii $r<300~\mu$m.

\section{Experimental results}
\label{section4}

When the viscous drop spreads on the substrate, we can clearly distinguish two regimes in the spreading. Fig. \ref{fig2}(a) shows the radius of the wetted area as a function of time for a drop of initial radius $R=0.5\pm 0.02~$mm and viscosity $\eta=220~$Pa.s. As expected from Tanner's law, the spreading is rather slow at late times ($t>10^{-2}$s), with contact line speeds smaller than $1~$mm.s$^{-1}$ in the range of our experiments. By contrast, during the first stages the dynamics is much quicker, with a contact line speed measured up to $\sim 0.5~$m.s$^{-1}$. The cross over in between these two regimes occurs at time $t\simeq 10^{-3}~$s after contact, for a radius $r\simeq 0.4~$mm. These features (two regimes and a cross-over) can be observed in all our experiments. The precise time and radius at which the cross-over occurs depends on the experimental parameters, namely the droplet initial radius $R$, the liquid viscosity $\eta$ and the surface wettability. In this section, we will systematically investigate the role of these parameters. In addition, the spatio-temporal resolution of our setup will enable us to characterize the spreading dynamics for the initial stages after contact. 

\subsection{Radius dependence}

\begin{figure}[t]
\includegraphics[width=1 \columnwidth]{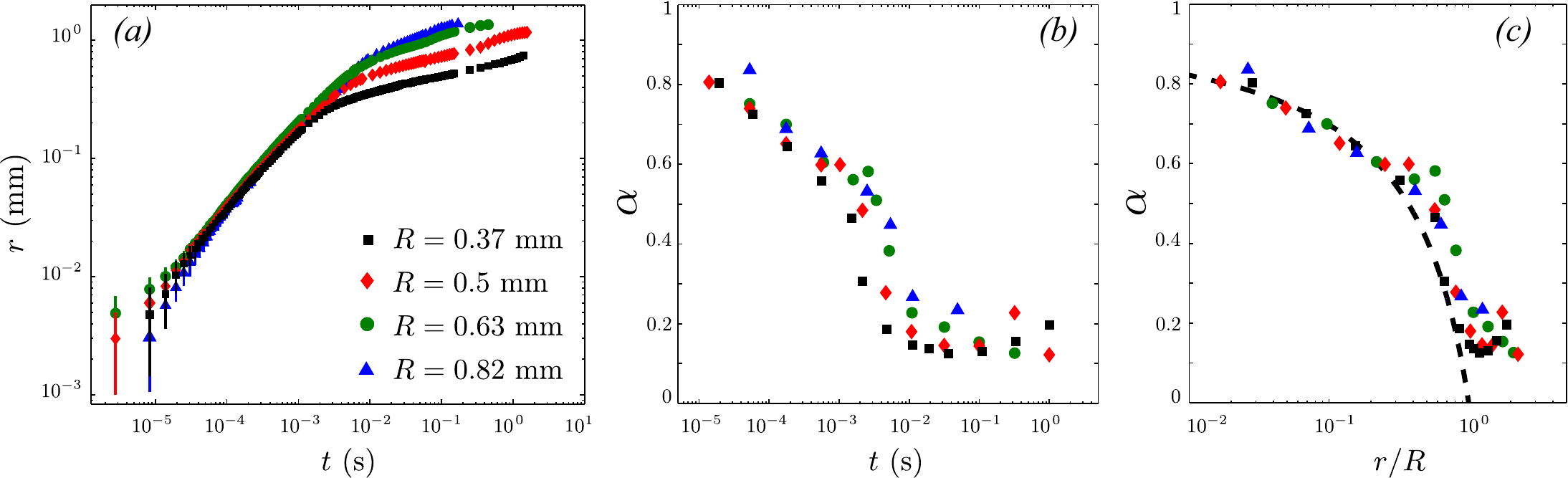}
\caption{\label{fig4} (a) Evolution of the radius $r$ of the wetted area as a function of time $t$ for drops with viscosity $\eta=50~$ mPa.s and initial radius $R=0.37~$mm (Black squares), $R=0.5~$mm (Red diamonds), $R=0.63~$mm (Green circles) and $R=0.82~$mm (Blue triangles). (b) Evolution of the apparent exponent $\alpha=\frac{d \ln{r}}{d \ln{t}}$ as a function of time for the same drops as in (a). (c) Evolution of $\alpha$ as a function of the adimensional radius $r/R$ of the wetted area for the same drops as in (a). The dashed line is the prediction from Eq. (\ref{eq_alpha}).}
\end{figure}

We will first look at the effect of the initial drop radius $R$ for a given liquid viscosity. Figure \ref{fig4}(a) shows the evolution of the radius of the wetted area as a function of time for drops of viscosity $\eta=50~$mPa.s for four different initial drop radii $R$. At short times, $r$ follows quantitatively the same dynamics for all drop radii $R$, and all the curves collapse on each other. During this first regime of spreading, $r(t)$ does not seem to depend on the drop size. Around a cross-over time $t_C$ and a cross-over radius $r_C$, each of the experimental curves undergoes a transition and present a sharp change to a slower dynamics. The values of $t_C$ and $r_C$ depend on the initial radius $R$: the cross-over occurs later for larger values of $R$. 

A common approach to characterize the spreading dynamics is to look for a power-law dynamics, $r\sim t^{\alpha}$. To investigate whether our experiments indeed display power-law spreading regimes, we define the apparent exponent 
\begin{equation}\label{eq:alpha}
\alpha=\frac{d \ln{r}}{d \ln{t}}~,
\end{equation}
and determine it from our experimental data by computing twice the value of $\alpha$ per decade of time. Figure \ref{fig4}(b) presents the evolution of $\alpha$ as a function of time $t$ for four different values of the drop radius $R$. At short times, there is a fast dynamics with an apparent exponent $0.5<\alpha<1$.  

Note that the values at early times are significantly larger than the slope 0.5 that was measured in recent experiments \cite{Carlson:2012}. This difference can be explained from the improved spatio-temporal resolution of our bottom view setup, giving access to earlier dynamics. In all cases, $\alpha$ slowly decreases in the first regime and abruptly decreases at a time $t_C$.  Later on, the apparent exponent reaches a plateau with a value $0.1<\alpha<0.2$, which is consistent with Tanner's law (Eq. (\ref{Tanner})). 

Interestingly, the early stage is not characterized by a single exponent $\alpha$: the apparent exponent is slowly decreasing with time. We emphasize that the cross-over to Tanner's regime is much more abrupt than this slow decrease at early times. Hence, we infer that the slowly decreasing $\alpha$ is not simply a cross-over effect, but an intrinsic property of the initial spreading dynamics. This becomes even more apparent in fig. \ref{fig4}(c), where we plot $\alpha$ as a function of the dimensionless radius $r/R$. For all initial radii $R$, we observe a collapse of the curves. The sharp transition to Tanner's regime occurs for a dimensionless radius $r_c/R\simeq0.8$. This value of $r_c/R$ can be understood from the geometry of the spreading drop (see fig. \ref{fig1}). At short times, the wetted area is located under a nearly spherical drop whereas, at late times, we recover a spherical cap of liquid that slowly spreads. The cross-over in-between these two geometrical configurations occurs when the radius of the wetted area is of the same order as the initial drop radius $R$, in good agreement with the observed value $r_c/R\sim1$. 

Let us now compare the short-time dynamics with the prediction provided by viscous drop coalescence \cite{Eggers:1999}. Equation (\ref{eq_radius}) is not a pure power-law, since it displays a logarithmic correction. Applying the definition Eq. (\ref{eq:alpha}), the effective exponent for coalescence can be derived as 

\begin{equation}
\alpha=\frac{\ln{\frac{r}{R}}}{\ln{\frac{r}{R}}-1}.
\label{eq_alpha}
\end{equation}
The dashed line in fig. \ref{fig4}(c) shows this prediction, which is in quantitative agreement with the experimental observations at early times. In particular, it captures the slow decrease of the exponent. This is a first indication that the initial dynamics of viscous drop spreading is similar to the coalescence of spherical drops.

\subsection{Wettability dependence}

\begin{figure}[t]
\begin{centering}
\includegraphics[width=1 \columnwidth]{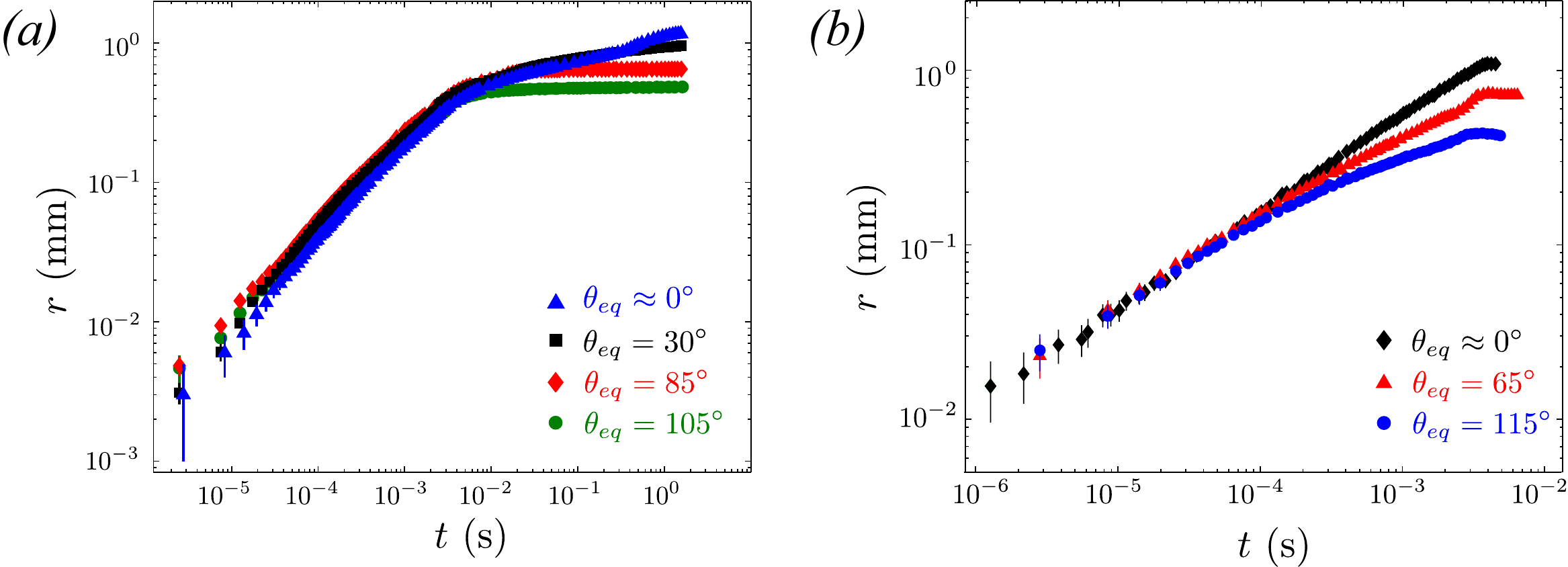}
\caption{\label{fig6} (a) Evolution of  $r$ as a function of $t$ for droplets with viscosity $\eta=50~$mPa.s and initial radius $R=0.5~$mm for 4 different substrates. Blue triangles: cleaned glass ($\theta_{eq}\simeq 0^{\circ}$), black squares: raw glass ($\theta_{eq}=30^{\circ}$), red diamonds: fluorinated coating ($\theta_{eq}=85^{\circ}$) and green circles: 3M coating ($\theta_{eq}=105^{\circ}$). The initial dynamics is identical for all $\theta_{eq}$ within experimental error. (b) Evolution of $r$ as a function of time for pure water drops and 3 different substrates with equilibrium contact angles $\theta_{eq}= 3^{\circ}, 65^{\circ}$ and $120^{\circ}$ \cite{Winkels:2012}.}
\end{centering}
\end{figure}

In order to check the wettability dependence of spreading, we performed experiments on three other substrates presenting various equilibrium contact angles $\theta_{eq}$. We used untreated glass that gives $\theta_{eq}\simeq30^{\circ}$. The two other substrates are glass slides covered by fluorinated coatings that increase the hydrophobicity of the substrate, limiting the spreading of the drop and imposing a finite equilibrium radius. The first one gives $\theta_{eq}\simeq85^{\circ}$, and the second one (Novec 1700, 3M) leads to $\theta_{eq}\simeq105^{\circ}$ . Fig. \ref{fig6} presents $r$ as a function of time for these four different substrate wettabilities. All the curves follow the same dynamics in the first regime, subsequently the spreading slows down and eventually stops when the drop reaches its equilibrium shape. The wettability has indeed almost no effect on the first regime, and it is only later that a deviation can be observed. 

\begin{figure}[p]
\includegraphics[width=.74 \columnwidth]{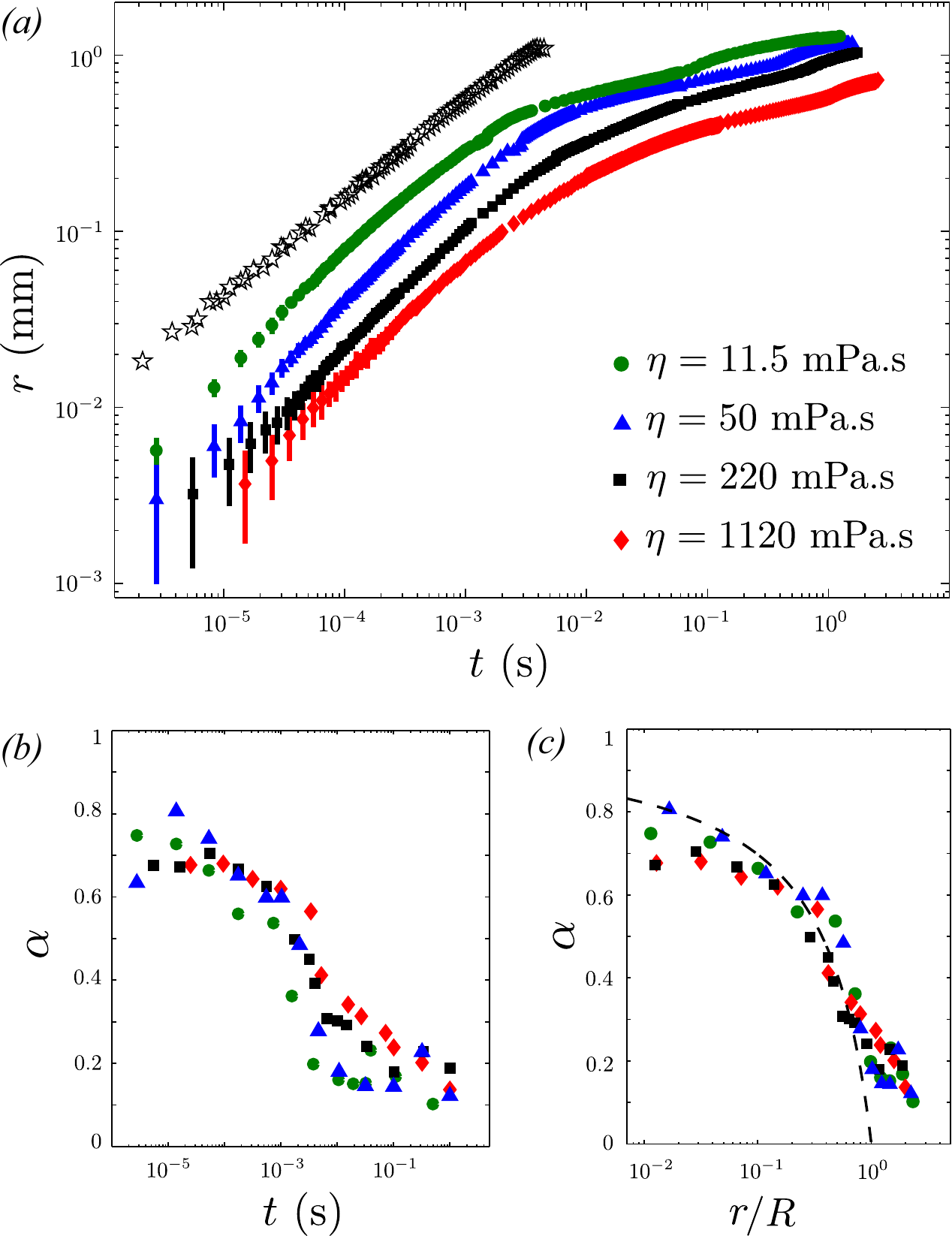}
\caption{\label{fig5} (a) Evolution of the radius $r$ of the wetted area as a function of time $t$ for drops in complete wetting with initial radius $R=0.5~$mm with viscosity $\eta=11.5~$ mPa.s (Green circles), $\eta=50~$ mPa.s (Blue triangles), $\eta=220~$ mPa.s (Black squares) and $\eta=1120~$ mPa.s (Red diamonds). The open stars are data obtained for pure water \cite{Winkels:2012}. (b) Evolution of the measured apparent exponent $\alpha$ as a function of the time for the same data as in (a). (c) Evolution of $\alpha$ as a function of $r/R$ for the same data as in (a). The dashed line is the prediction from Eq. (\ref{eq_alpha}).}
\end{figure}

These observations are consistent with the hypothesis that the initial spreading dynamics is similar to coalescence: the independence of wettability suggests that the solid wall is unimportant at the initial stages. Let us emphasize that the same conclusion was previously drawn for the case of pure water, which have a much lower viscosity \cite{Biance:2004, Winkels:2012}. The spreading experiments for water are reported in fig. \ref{fig6}(b). For three different wettabilities, the spreading dynamics is identical at short times \cite{Winkels:2012}. In this case, the dynamics is characterized by a well-defined exponent $\alpha=1/2$. This agrees with the law for inertially dominated coalescence that predicts
\begin{equation}
\frac{r}{R}=D_0 \left(\frac{t}{\sqrt(\rho R^{3}/\gamma)}\right)^{\frac{1}{2}}~.
\end{equation}
The value of the prefactor $D_0=1.14$ has been determined experimentally \cite{Aarts:2005}.  For water drops spreading on a substrate \cite{Winkels:2012}, $D_0=1.2\pm 0.1$, in excellent agreement with the coalescence case.

\subsection{Viscosity dependence}

We now investigate how viscosity affects the spreading dynamics. Fig. \ref{fig5}(a) shows the evolution of $r(t)$ for drops with initial radius $R=0.5~$mm and five different viscosities. The closed symbols are experiments for water-glycerine mixtures, while for completeness we also report the inertia-dominated case of pure water (open symbols) \cite{Winkels:2012}). For the viscous drops, all curves again exhibit the same qualitative behavior, characterized by two regimes. As expected, the dynamics is much slower for the drops of higher viscosity. Once more, there is no clear spreading exponent at early times. This can be seen from fig. \ref{fig5}(b) showing the evolution of $\alpha$ as a function of time $t$. The apparent exponent slowly decreases during the first regime, and reaches a plateau around $\alpha \simeq 0.15$. The time for the cross-over depends on the viscosity. The cross-over appears at earlier times and is sharper for lower viscosities. However, this effect of viscosity can be eliminated when plotting $\alpha$ as a function of the dimensionless radius $r/R$ (fig. \ref{fig5}(c)). We observe a collapse of the data onto the a master curve, again in quantitative agreement with the prediction from Eq. (\ref{eq_alpha}). This strongly suggests that the dynamics of viscous drop spreading depends on the geometry (given by the dimensionless radius $r/R$) at short times, in a manner consistent with the coalescence of viscous drops. 

\begin{figure}[t]
\includegraphics[width=1 \columnwidth]{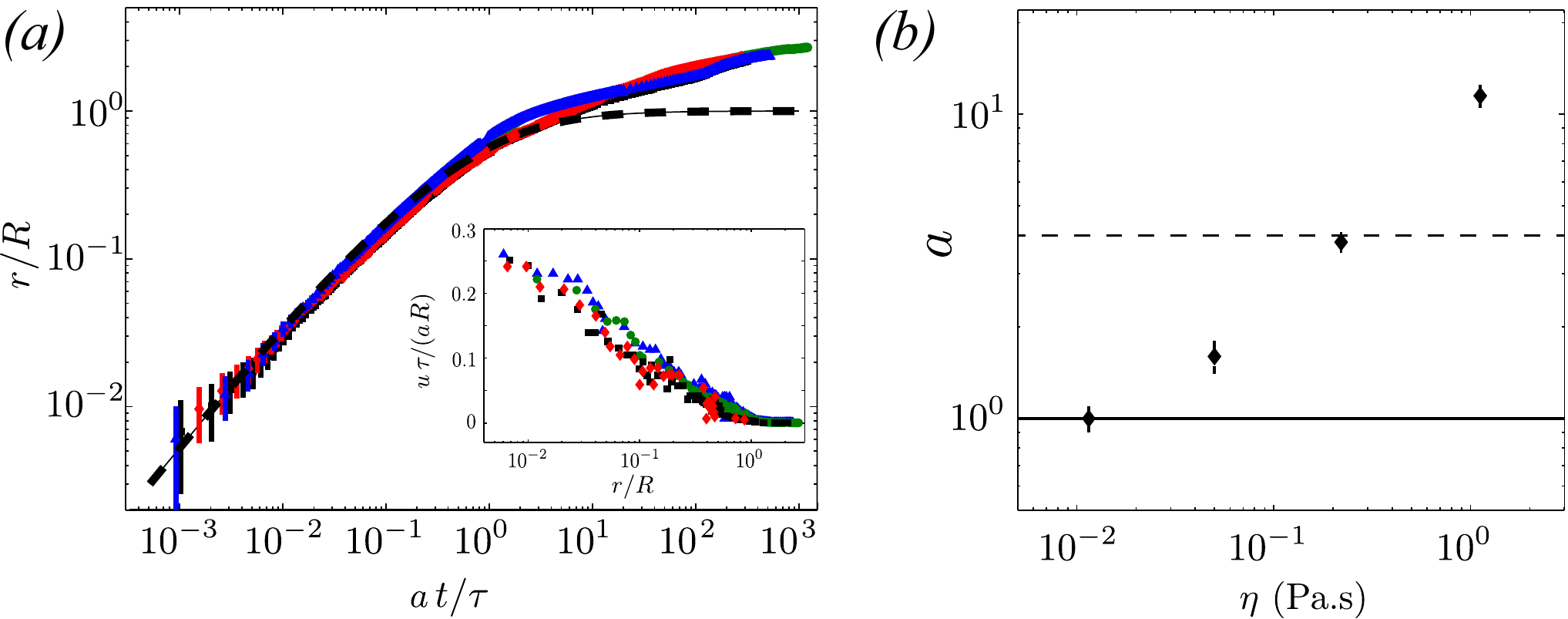}
\caption{\label{fig7} (a) Dimensionless radius $r/R$ as a function of dimensionless time $a t/\tau$, with $\tau= 4 \pi \eta R/\gamma$ for data from fig. \ref{fig5}. The dashed line is the theoretical prediction from Eq. (\ref{eq_radius}). Inset: dimensionless contact line speed $u \tau/(aR)$ as a function of the adimensionnal radius $r/R$. The collapse agree with the logarithmic decay prediction by  Eq. (\ref{eq_speed}). (b) Values of the fitted prefactor $a$ for different liquid viscosities $\eta$.}
\end{figure}

Finally, based on the prediction Eq.~(\ref{eq_radius}), we scale the data to the form
\begin{equation}
\frac{r}{R} =  - a \frac{t}{\tau} \ln \frac{r}{R}, \quad {\rm with} \quad \tau = \frac{4\pi \eta R}{\gamma},
\label{eq_datafitting}
\end{equation}
where $a$ is used as an adjustable parameter. In coalescence of spherical drops, the parameter $a=4$ in the case where the gas viscosity can be neglected, while it is predicted that $a=1$ for the case of non-neglible gas viscosity \cite{Eggers:1999}. Figure \ref{fig7}(a) reveals that an excellent collapse to this form can be achieved for the initial spreading. The prefactors $a$ for the various viscosities are reported in Figure \ref{fig7}(b). Interestingly, the value of $a$ is not universal, but increases significantly with $\eta$. Qualitatively, this is similar to viscous coalescence, for which $a$ is expected to increase for increasing liquid viscosity. However, the effect is stronger than expected from theory, which predicts that $a$ should remain within the boundaries from 1 to 4. 

Finally, it should be noted that Eq.~(\ref{eq_radius}) relies on the fact that the spreading speed $u=dr/dt$ is imposed by the Stokeslet solution \cite{Eggers:1999}, and is of the form
\begin{equation}\label{eq:bah}
u = -~\frac{a R}{\tau} ~ \ln{\frac{r}{R}}.    
\end{equation}
Hence, the spreading velocity should decrease logarithmically as $r/R$ increases. We indeed find such a decrease in spreading velocity, as can be seen from the inset on fig. \ref{fig7}(a). The collapse was obtained with the same values of the fitting parameter $a$.  Let us note that the dynamics of Eqs.~(\ref{eq_radius}) and (\ref{eq:bah}) are equivalent only asymptotically, and that for finite ratio $r/R$ there are small (logarithmic) differences. In practice, this leads to $30\%$ underestimation of the velocities with respect to Eq.~(\ref{eq:bah}). 

\section{Discussion}

\label{section5}

The initial spreading of low-viscosity drops exhibits a well-defined exponent $r \sim t^{1/2}$ over several decades \cite{Biance:2004,Winkels:2012}. In this case, the spreading was found identical to that of the dynamics of coalescence of two low-viscosity drops \cite{Aarts:2005}, which exhibits the same exponent and the same prefactor. In this paper we have shown that for viscous drops spreading on a substrate, the evolution of the wetted area is not characterized by a simple power-law growth, as was recently proposed \cite{Carlson:2012}. During the initial spreading stages, the apparent exponent $\alpha = d\ln r/d\ln t$ is observed to slowly decrease from about 0.8 to 0.5. Once more, it turned out fruitful to compare the spreading dynamics to the coalescence of viscous drops. Namely, the coalescence is characterized by a slowly decreasing velocity \cite{Hopper:1993}, and such a dynamics agrees quantitively with all our spreading experiments. The analogy with coalescence of freely suspended drops suggests that the presence of the wall has no significant influence on the flow. This picture is confirmed by our observation that the initial spreading is completely independent of the wettability of the substrate. This is particularly surprising since the flow develops singular viscous stresses near the contact line \cite{Huh:1971}. It thus appears that the singularity due to the initial contact, when the spherical drop first touches the flat substrate, completely dominates over the contact line singularity. It is only later, when the drop enters in Tanner's regime, that  the contact line dissipation controls the spreading dynamics.

While the data for different viscosities can be collapsed onto the form predicted for coalescence (see fig. \ref{fig7}(a)), the typical spreading velocity does not simply scale as $\sim 1/\eta$. This can be seen from the prefactor $a$, which still displays a dependence on viscosity $\eta$ (see fig. \ref{fig7}(b)). A similar observation was made recently by Carlson et al. \cite{Carlson:2012}. They noticed that the spreading data cannot be collapsed using a simple viscous scaling, and introduced additional friction at the contact line to explain this result. Here we suggest that such a dependence can (at least partly) be explained from the influence of the outer fluid. The effect of the gas was previously taken into account for the coalescence problem \cite{Eggers:1999}, indeed giving rise to non-universal values of the prefactor $a$. However, the values of $a$ necessary to fit our data explore a wider range than predicted for coalescence -- in particular in the large viscosity regime where the outer fluid should be negligible. We nevertheless find prefactors that are comparable to other coalescence experiments \cite{Paulsen:2011}, so that experimentally the ``spreading--coalescence'' analogy might still be fully quantitative. Interestingly, the strong influence of the gas was also emphasized for air entrainment by advancing contact lines \cite{Marchand:2012}. In that case the speed dependence on liquid viscosity was found much weaker than $1/\eta$, in a way comparable to the present experiments. Similarly, impact experiments were collapsed using a $1/\sqrt{\eta}$ speed-dependence \cite{deRuiter:2012}. It thus seems that the confinement of the air near the contact line has a major effect, but the details have remained unresolved.

The key open question is why the spreading--coalescence analogy provides such an accurate description of the experiment. Even though the flow during the coalescence of two spherical drops exhibits a mirror-plane, there is still a substantial flow along this plane of symmetry. For spreading, this plane coincides with the wall and one would have expected that the no-slip condition strongly impedes the motion of the contact line. A possible interpretation could be that slip is enhanced due to the singular geometry of the initial contact. 

Acknowledgments -- We gratefully acknowledge Jens Eggers and Joost Weijs for fruitful discussions. This work was funded by VIDI grant N$^\circ$ 11304 and is part of the research program 'Contact Line Control during Wetting and Dewetting' (CLC) of the 'Stichting voor Fundamenteel Onderzoek der Materie (FOM)', which is financially supported by the 'Nederlandse Organisatie voor Wetenschappelijk Onderzoek (NWO)'. The CLC programme is co-financed by ASML and Océ.

\bibliography{bibliography.bib}

\end{document}